\documentclass{article}

\usepackage{multirow}
\usepackage{multicol}
\usepackage[flushleft]{threeparttable}

\usepackage{PRIMEarxiv}

\usepackage{natbib}
\usepackage{xcolor}

\usepackage[utf8]{inputenc} % allow utf-8 input
\usepackage[T1]{fontenc}    % use 8-bit T1 fonts
\usepackage{hyperref}       % hyperlinks
\usepackage{url}            % simple URL typesetting
\usepackage{hyphenat}      % for \showhyphens command
\usepackage{booktabs}       % professional-quality tables
\usepackage{amsfonts}       % blackboard math symbols
\usepackage{nicefrac}       % compact symbols for 1/2, etc.
\usepackage{microtype}      % microtypography
\usepackage{lipsum}
\usepackage{fancyhdr}       % header
\usepackage{graphicx}       % graphics
\DeclareMathSymbol{\mhyphen}{\mathord}{AMSa}{"39}

\usepackage{array}
\usepackage{float}
\newcolumntype{P}[1]{>{\centering\arraybackslash}p{#1}}

\title{CiteBART: Learning to Generate Citations for Local Citation Recommendation
}

\author{
  Ege Yiğit ÇELİK \\
  Computer Engineering Department \\
  İzmir Institute of Technology \\
  İzmir, Turkey \\
  \texttt{egecelik@iyte.edu.tr} \\
   \And
  Selma TEKİR \\
  Computer Engineering Department \\
  İzmir Institute of Technology \\
  İzmir, Turkey \\
  \texttt{selmatekir@iyte.edu.tr} \\
}

\begin{document}
\maketitle

\begin{abstract}
Local citation recommendation (LCR) suggests a set of papers for a citation placeholder within a given context. The task has evolved as generative approaches have become more promising than the traditional pre-fetch and re-rank-based state-of-the-art approaches. This paper introduces citation-specific pre-training within an encoder-decoder architecture, where author-date citation tokens are masked to learn to reconstruct them to fulfill LCR. There are two variants for this pre-training. In the local context-only base scheme (CiteBART-Base), the citation token in a local context is masked to learn to predict the citation. The global version (CiteBART-Global) extends the local context with the citing paper's title and abstract to enrich the learning signal. CiteBART-Global achieves state-of-the-art performance on LCR benchmarks except for the FullTextPeerRead dataset, which is quite small to see the advantage of generative pre-training. The effect is significant in the larger benchmarks, e.g., Refseer and ArXiv., with the Refseer benchmark-trained model emerging as the best-performing model. We perform comprehensive experiments, including an ablation study, a qualitative analysis, and a taxonomy of hallucinations with detailed statistics. Our analyses confirm that CiteBART-Global has a cross-dataset generalization capability; the macro hallucination rate (MaHR) at the top-3 predictions is 4\%, and when the ground-truth is in the top-k prediction list, the hallucination tendency in the other predictions drops significantly.
We publicly share our code\footnote{\url{https://github.com/eyclk/CitationRecommendation}}, base datasets\footnote{\url{https://drive.google.com/drive/folders/1WlqlTkSj8LwihbrQvBX5F9_0uZAGGhiE?usp=drive_link}}, global datasets\footnote{\url{https://drive.google.com/drive/folders/1JH34nEXt8_p-0P9A--aQHK4yBXQfJe4v?usp=drive_link}}, and pre-trained models\footnote{\url{https://drive.google.com/drive/u/2/folders/1OBg6W3kQw4VWPMfrXEPxN8LzTopR1jak}} to support reproducibility. 

\end{abstract}

% keywords can be removed

% ********************************************************************************************************
\section{Introduction}

Citations are essential building blocks in scientific writing. Their accurate placements indicate quality, as one should know the literature to claim contributions and put the current study in the context of the existing work from different aspects, such as background information, method, and result comparison \citep{cohan-etal-2019-structural}.

%The first citation-related task in natural language processing (NLP) has been citation impact prediction, where a paper's future scientific impact is predicted on the basis of the number of times a paper gets cited after publication \citep{gehrke2003}. Unlike the first approaches that relied on paper metadata and abstract, the recent work (\citet{van-dongen-etal-2020-schubert}, \citet{huang2022}) exploit the whole content of scientific papers to achieve the goal.

Citation prediction is defined as a two-step process where the former focuses on where in the sentence to place the citation \citep{buscaldi-etal-2024-generative-cite}, while the latter (citation recommendation) obtains a set of candidate papers once there is a specified citation placeholder in a given context. In this sense, citation recommendation serves as a citation suggestion mechanism. For a given scientific text, it can suggest additional papers on a similar topic. These suggestions can be considered additional reading material alongside the targeted paper, corresponding to the ground-truth citation.

There are two levels of citation recommendation: the first, whom to cite, and the second, whom to cite in what context. The former is global citation recommendation, traditionally performed based on paper metadata such as author names, paper titles, abstracts, conference venues, publisher information, etc. Recently, custom citation-aware language models (SciBERT \citep{beltagy-etal-2019-scibert}, SPECTER \citep{cohan-etal-2020-specter}) learn good citation-aware embeddings for full papers to perform well in this task. The latter task is local citation recommendation (LCR), aiming to determine the target paper for a citation placeholder. %Additionally, the local citation contexts can be leveraged for citation impact prediction. How a paper frames its work through citations is predictive of the citation count it will receive \citep{jurgens-etal-2018-measuring}. 

%Language model pre-training based on Transformers \citep{vaswani2017attention} provided new state-of-the-art performances in many downstream tasks. Masked language modeling (MLM) is the primary learning strategy behind BERT \citep{devlin-etal-2019-bert} and its variants (RoBERTa \citep{liu2019roberta} etc.). 

\begin{figure*}[th]
   \centering
   \includegraphics[width=0.88\linewidth]{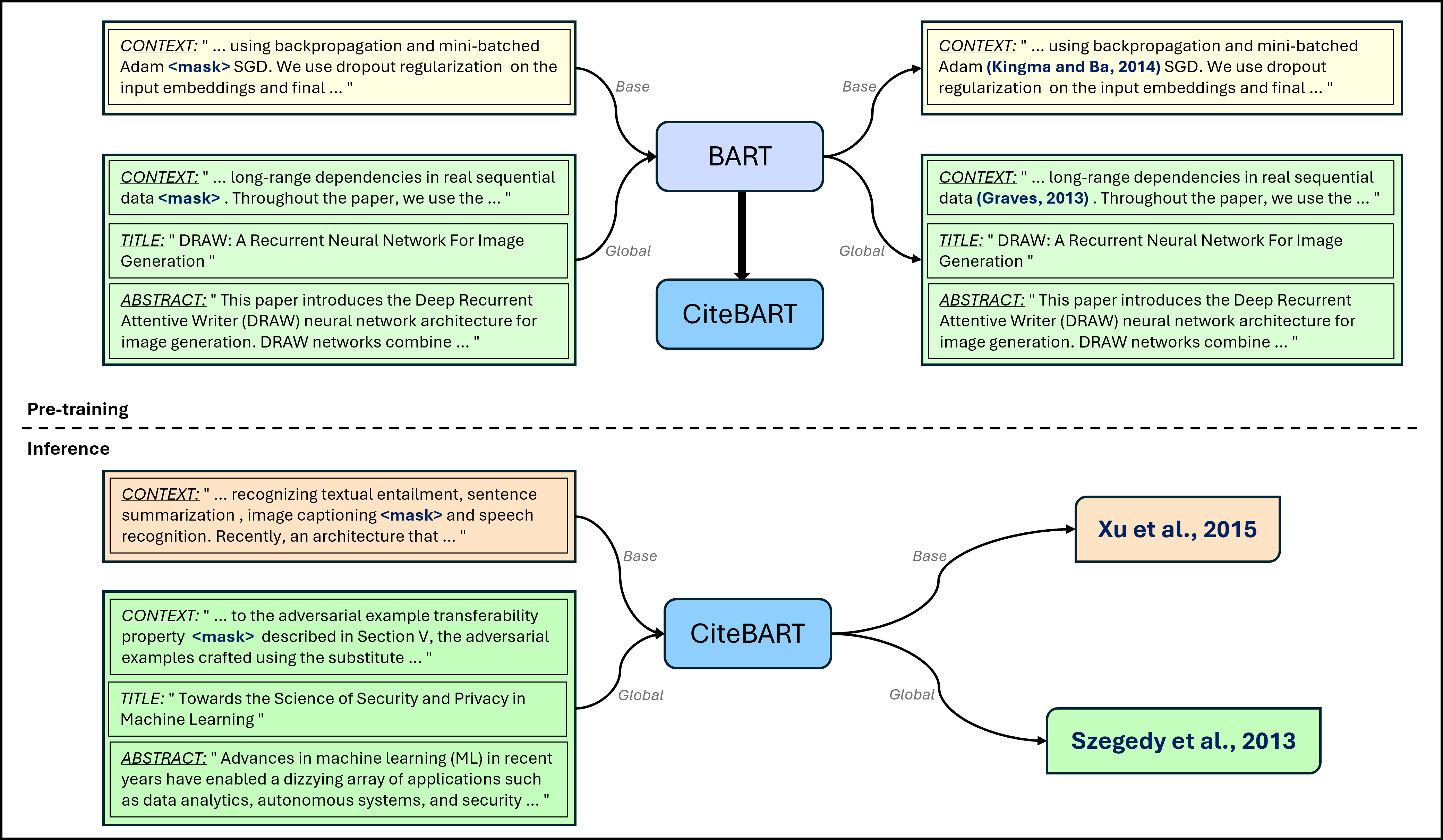}
   \caption{CiteBART workflow. The yellow and green examples represent the workings of CiteBART-Base and CiteBART-Global, respectively. During inference, the expected outputs are in the author–date citation format, unlike the pre-training stage.}
   \label{fig:citebart-architecture}
\end{figure*}

LCR has been addressed in a few works. BERT-GCN \citep{jeong-etal-2020-bertgcn} utilizes a feedforward neural network to combine local citation context representations using BERT with citation encodings through Graph Convolutional Neural Networks (GCN). The most recent solutions to the problem adopt a two-step process that consists of pre-fetching and re-ranking. DualEnh \citep{medic-snajder-2020-improved} enhances a local citation context with the citing article's title and abstract and uses this enhanced context as the query vector to retrieve the most similar candidate articles using their titles and abstracts. It performs this ranking through BiLSTM representations of inputs with attention layers on top.
On the other hand, HAtten \citep{gu2022hatten} initially pre-fetches a set of papers using the nearest neighbor search between local citation context extended with the citing paper's title and abstract (query text as a whole) and the title and abstracts from a given pool of papers. Afterward, it re-ranks the selected candidate papers using a fine-tuned SciBERT \citep{beltagy-etal-2019-scibert} model where the input is the query text concatenated with a candidate paper's title and abstract. SymTax \citep{goyal-etal-2024-symtax} improves upon HAtten by introducing an additional Enricher module and reranking candidate papers using taxonomical relationships along with contexts.

The existing LCR works are not built upon Transformers but benefit from it indirectly, such as re-ranking the results (using fine-tuned SciBERT). Distinctively, we propose CiteBART, a custom pre-training approach based on the Transformer architecture. We mask citation tokens in the local contexts to learn to reconstruct them effectively during pre-training.

\citet{fierro-etal-2024-learning} support information-seeking using query-focused summarization, responding to user queries by answers with source attributions. Attributions are in the form of in-line references to the passages. In a similar direction, the ALCE benchmark \citep{gao-etal-2023-enabling} collects a diverse set of questions and retrieved passages to support answer generation with appropriate citations. As these models exhibit citation generation abilities, they are similar to CiteBART. However, CiteBART aims to fill in a passage with citations instead of targeting retrieval-based summaries with citations. In ALCE, there is a closed-book configuration where the model does not access any retrieved document to generate answers to a user query but is still different as the focus is answering a query instead of filling in the citation placeholder.

%The query-focused summarization datasets consist of queries, passages, and target summaries. The task is to generate a summary with attributions given the query and passages. 

\begin{table*}[htb]
 \caption{An example for input and target formats for evaluation with CiteBART. Due to space constraints, we present the contexts and abstracts in an abbreviated form.}
  \centering
  \scalebox{0.84}{
  \begin{tabular}{lp{132mm}p{35mm}}
    \toprule
    Strategy & Input & Target \\
    \midrule
    Base & … error rate of 5.8\% and a word error rate of 28.7\%, which are on par with previous reported results \textbf{<mask>} . Unlike prior work, we do not use a language model during decoding and … & \textbf{Yao and Zweig, 2015} \\
    \midrule
    Global & … error rate of 5.8\% and a word error rate of 28.7\%, which are on par with previous reported results \textbf{<mask>} . Unlike prior work, we do not use a language model during decoding and … \textit{</s>} Deep Voice: Real-time Neural Text-to-Speech \textit{</s>} We present Deep Voice, a production-quality text-to-speech system constructed entirely from deep neural … & \textbf{Yao and Zweig, 2015}  \\
    \bottomrule
  \end{tabular}}
  \label{tab:input-target-formats-bart}
\end{table*}

%BART \citep{lewis-etal-2020-bart} is a generative model that is pre-trained using a seq-to-seq objective while employing a mask-filling mechanism similar to that of BERT. Its MLM objective allows for a random subset of the tokens to be masked and multiple tokens to be predicted. Due to this capability, BART is particularly well-suited to our approach, aiming to generate complex parenthetical author-date citations without requiring further architectural modifications. Furthermore, BART's encoder, like BERT, effectively captures contextual information bidirectionally. Additionally, BART's generative nature permits the adjustment of generation parameters, providing greater control over the evaluation of the generated citations without further pre-training. 

In our approach, the base scheme (CiteBART-Base) learns through the masked citation context. In a second technique (CiteBART-Global), we extend the masked context with the citing paper's global information, e.g., title and abstract (Table \ref{tab:input-target-formats-bart}). Inspiring from pre-training under the REALM framework \citep{guu2020realm}, we append this global information to the local context, allowing backpropagation through the global information to learn associations with the pool of papers from the corpus.

CiteBART achieves superior performance without relying on a pre-fetch and re-rank pipeline. It is an end-to-end learning system. On the other hand, a pre-fetch and re-rank pipeline, such as HAtten, utilizes the citing papers' titles and abstracts with the local contexts to form the query encoding and the titles and abstracts of cited papers for the candidate papers' representations. Thus, it exploits the titles and abstracts of papers from the test set to determine the cited papers for a citation placeholder. On the contrary, we do not use the global information (titles and abstracts) of target papers to make the recommendation. CiteBART-Global learns solely from the relation of citing papers' global information with local citation contexts. The underlying assumption is one can find out the cited papers from the enhanced citation contexts, citation contexts that are concatenated with citing papers' titles and abstracts. In the test phase, we feed these enhanced contexts to predict the target papers to be cited. %Furthermore, CiteBART can be fine-tuned for any downstream task. 

CiteBART presents a novel perspective to LCR. It achieves superior performance without relying on a pre-fetch and re-rank pipeline. It is an end-to-end learning system. Unlike previous works, we do not exploit the global information (titles and abstracts) of the target papers to make the recommendation. CiteBART-Global learns solely from the relation of citing papers' global information with local citation contexts.

% We share our code\footnote{\url{https://github.com/eyclk/CitationRecommendation}}, base datasets\footnote{\url{https://drive.google.com/drive/folders/1WlqlTkSj8LwihbrQvBX5F9_0uZAGGhiE?usp=drive_link}}, and global datasets \footnote{\url{https://drive.google.com/drive/folders/1JH34nEXt8_p-0P9A--aQHK4yBXQfJe4v?usp=drive_link}} publicly for supporting reproducibility.

We summarize our contributions as follows:
\begin{itemize}
    \item We propose an end-to-end learning system, CiteBART, with custom citation masking for LCR.
    \item CiteBART-Global achieves state-of-the-art performance on LCR benchmarks except for the FullTextPeerRead dataset, which is quite small to see the advantage of generative pre-training. The effect is significant in the larger benchmarks, e.g., Refseer and ArXiv. CiteBART-Base is still a strong baseline.
    \item We provide a qualitative analysis to gain insight into the working of the approach, including the cross-dataset generalization capability.  % zero-shot
    \item We provide a taxonomy of hallucinated citations and report macro hallucination rates (MaHR) for them.
    \item Our ablation study confirms the central role of local citation contexts in the learning process. It also shows the effectiveness of the Global training scheme over Base.
\end{itemize}

% ***********************************************************************************************************

\section{Related Work}
\label{sec:related_work}

BERT \citep{devlin-etal-2019-bert} is an encoder-only pretraining model that adopts the Masked Language Modeling (MLM) objective. MLM masks tokens in a uniformly random fashion and predicts them, allowing the generation of learning signals bidirectionally. Some BERT variants were released to meet the requirements for masking a group of tokens. SpanBERT \citep{joshi2020} builds on this objective by masking random contiguous text spans. In the same direction, PMI-Masking \citep{levine2021} masks word n-grams based on their PMI (Pointwise Mutual Information) scores. Pretraining encoder decoders, e.g., BART \citep{lewis-etal-2020-bart}, combine the strengths of bidirectional learning of encoders with the autoregressive nature of decoders, capturing the local patterns of tokens within their generative capabilities.

The first citation-related task in natural language processing (NLP) has been citation impact prediction, where a paper's future scientific impact is predicted on the basis of the number of times a paper gets cited after publication \citep{gehrke2003}. Unlike the first approaches that relied on paper metadata and abstract, the recent work (\citet{van-dongen-etal-2020-schubert}, \citet{huang2022}) exploit the whole content of scientific papers to achieve the goal. \citet{brody2006} aim to predict the future citations of a paper using web usage statistics. NNCP \citep{abrishami2019} uses a SimpleRNN model to predict long-term citations using short-term citations. \citet{bai2019} propose the Paper Potential Index based on a combination of manually acquired features. SChuBERT \citep{van-dongen-etal-2020-schubert} leverages the entire contents of papers to accomplish the task. FGCCP \citep{huang2022} performs a fine-grained analysis to attribute citation frequencies to individual parts of papers.

\citet{Yu2012CitationPI} learn citation relations through a meta path-based approach. Their approach combines authorship metadata with discriminative term features to calculate citation probabilities on the DBLP network. \citet{tanner-charniak-2015-hybrid} combine LDA-Bayes with metadata features under a logistic regression classifier to recommend citations. 

Similar to citation recommendation, the recent work of \citet{luo-etal-2023-prototype} predicts provisions of the U.S. Code by pretraining RoBERTa \citep{liu2019roberta} and LegalBERT \citep{chalkidis-etal-2020-legal} on the curated dataset (PACER \citep{luo-etal-2023-prototype}) of the US federal court documents where each provision source text is given with its associated target citation. SciBERT \citep{beltagy-etal-2019-scibert} performs pretraining exclusively on scientific texts to learn global representations for scientific papers. SPECTER \citep{cohan-etal-2020-specter} learns citation-aware global representations for scientific papers using a citation-based pretraining objective. SPECTER-produced representations introduced remarkable results in the paper classification and global citation recommendation tasks.

%In scientific document understanding, learning better representations for scientific papers has been a focus. In such an effort, SciBERT \citep{beltagy-etal-2019-scibert} performs pretraining exclusively on scientific texts. Specifically, it is pre-trained on a randomly sampled dataset of $1.14M$ from the Semantic Scholar database. It is built upon an in-domain vocabulary (SCIVOCAB), which brings about superior performance compared to BERT in downstream tasks that involve scientific data. 

%SPECTER \citep{cohan-etal-2020-specter} learns citation-aware global representations for scientific papers using a citation-based pretraining objective. Starting from the initial SciBERT weights, the system adopts a triplet loss function based on document similarities. The first component in this objective is the source document's similarity to one of its citations. In contrast, the second component is its similarity to a negative paper not cited by the source, and finally, there is an additional term of loss margin hyperparameter. SPECTER-produced representations introduced remarkable results in the paper classification and global citation recommendation tasks.

LCR has four benchmark datasets for evaluation. BERT-GCN \citep{jeong-etal-2020-bertgcn} introduced the FullTextPeerRead dataset, extended from the original PeerRead \citep{kang-etal-2018-dataset}. Throughout this paper, we refer to the FullTextPeerRead dataset as PeerRead for brevity. An additional dataset is ACL-ARC \citep{bird-etal-2008-acl-200-arc}, derived from the ACL Anthology Reference Corpus. We run our experiments on its ACL-200 subcategory, analogous to DualEnh \citep{medic-snajder-2020-improved} and HAtten \citep{gu2022hatten}. Finally, Refseer \citep{huang-etal-2015-refseer} and ArXiv \citep{gu2022hatten} are the largest benchmarks for this task.

BERT-GCN \citep{jeong-etal-2020-bertgcn} utilizes two encoders for citation recommendation. The first encoder generates local context embeddings using BERT, while the second one creates the graph embeddings of citation networks using a GCN model \citep{kipf2017semisupervised}. The approach combines these embeddings to produce representations for papers. It was evaluated exclusively on the PeerRead dataset.

DualEnh \citep{medic-snajder-2020-improved} trains a Bi-LSTM model to leverage similarity between a target paper and its candidate papers. The target paper provides a context with a citation placeholder, and the model utilizes the titles and abstracts of candidate papers to calculate their semantic similarity scores. The authors calculate semantic and bibliographic scores to acquire the final recommendation scores as a weighted average. The bibliographic score is acquired by utilizing metadata such as author names and citation counts. The authors performed their experiments on the ACL-200 and Refseer datasets.

HAtten \citep{gu2022hatten} uses a Hierarchical Attention Text Encoder and SciBERT-based Re-ranking scheme for LCR. It starts by pre-fetching potential candidate papers from a pool of citations. It accomplishes this filtering through a nearest neighborhood search between the local citation context plus the citing paper's title and abstract (query text as a whole) and the title and abstracts from candidate target papers. In the re-ranking phase, the authors assign scores to candidate papers using a SciBERT model with a classification layer on top. HAtten achieves state-of-the-art results on all of the benchmark datasets.

SymTax \citep{goyal-etal-2024-symtax} introduces a three-stage recommendation architecture for the LCR task, consisting of the Prefetcher, Enricher, and Reranker modules. Prefetcher is the same as HAtten's. Enricher leverages a pre-constructed citation network built from candidates to enhance their representation. Finally, Reranker combines a language model-based text relevance with a taxonomy relevance to yield a final recommendation. SymTax outperforms HAtten on the benchmark datasets.

Lastly, GM-s2orc-H \citep{buscaldi-etal-2024-generative-cite} proposes two approaches for predicting citation placeholders within a given context. Their first approach employs the GPT-2 model to determine whether a token could be part of a citation. The second approach performs a similar task using the BERT model, framing it as a Named Entity Recognition (NER) task. Their results confirm the superiority of the generative GPT-2 model over the second one. Although their results are not directly comparable to CiteBART due to differences in the task objectives, their findings highlight the advantages of generative models in citation-related tasks.

% Our work differs from the approaches \cite{fierro-etal-2024-learning} that strengthen information-seeking scenarios by generating supporting evidence in the form of in-line citations. These plan-based systems learn to relate questions with a set of passages during pre-training so that they can generate grounded answers with attributions to the retrieved passages for a given query.
% The related works are not directly relevant. They are abstractive/extractive summarization systems that perform retrieval based on a user query. They rely on a training dataset composed of queries, passages, blueprints, and summaries. We will state the differences on the camera-ready version to avoid misunderstandings.
% They focus on retrieval augmented generation with citations.
% Similarly, the ALCE benchmark \cite{gao-etal-2023-enabling} tests the citation abilities of retrieval augmented generation systems by providing the query and top-5 passages as input. When no passages are provided, the problem turns into finding the best passage for an input sentence (the closed book scenario), which is closest to the task of LCR. This scenario corresponds to the Vanilla LLM performance in LCR provided in Appendix D.

 % *******************************************************************************************************

\section{Methodology}
\label{sec:methodology}

We propose CiteBART, a novel pre-training strategy designed to predict citations within the contexts of scientific papers. We mask placeholder tokens, which replace ground-truth citations in the parenthetical author-date style, for the continual pre-training of a vanilla BART-base to generate the correct parenthetical author-date citation for a given context. CiteBART is trained on the benchmark datasets, learning to recommend citations during its generation process.

\subsection{Custom BART Pre-training for LCR}

BART \cite{lewis-etal-2020-bart} is a sequence-to-sequence model with an encoder and a decoder. It introduces a set of document corruption (denoising) schemes and then optimizes a reconstruction loss, the cross-entropy between the original document and the decoder's outputs. The denoising transformations that are applied to the encoder during pre-training are as follows: Random token masking (similar to BERT), token deletion, text infilling (span masking with span lengths drawn from a Poisson distribution ($\lambda = 3$)), sentence permutation, and document rotation with a randomly selected token leading the document.

We propose a citation learning strategy using BART. BART employs MLM similar to BERT. Additionally, to effectively reconstruct the masked contexts, it masks a span of $k$ tokens with a single mask. In return, it can predict multiple tokens for a single mask. Thus, CiteBART can generate complex parenthetical author-date citations after custom pre-training for citation tokens without requiring further architectural modifications.

We propose two training schemes for our approach: CiteBART-Base and CiteBART-Global (Figure \ref{fig:citebart-architecture}). In CiteBART-Base, the model gets the masked context with the ground-truth citation as input. This setting tests the model's performance in a local context-only situation (Table \ref{tab:input-target-formats-bart}). With the underlying idea that good citation recommendation requires relating local citation contexts with the citing papers' global information, such as titles and abstracts, we devised an innovative way to accomplish it. Inspiring from pre-training under the REALM framework \citep{guu2020realm}, in CiteBART-Global, we append the citing paper's title and abstract to the local context, allowing backpropagation through the global information that considers the pool of papers from the corpus. Specifically, we used the "$<$/s$>$" token designated by the pre-trained BART-base model as the separator.

\begin{table}[ht]
 \caption{Statistics of LCR benchmarks.}
  \centering
  \scalebox{0.95}{
  \begin{tabular}{lp{17mm}p{17mm}p{14mm}p{17mm}}
    \toprule
    Dataset Name & ACL-200 & PeerRead & RefSeer & Arxiv \\
    \midrule
    Train Size  & 30,390 & 9,363 & 3,521,582 & 2,988,030  \\
    Validation Size  & 9,381 & 492 & 124,911 & 112,779  \\
    Test Size  & 9,585 & 6,184 & 126,593 & 104,401  \\
    \# of Papers & 19,776 & 4,837 & 624,957 & 1,661,201  \\
    Publication Years  & 2009-2015 & 2007-2017 &   -2014 & 1991-2020  \\
    \bottomrule
  \end{tabular}}
  \label{tab:original-dataset-info}
\end{table}

\subsection{Dataset Preprocessing}

We conduct our experiments on the existing citation recommendation benchmarks of ACL-200, PeerRead, RefSeer, and Arxiv. Table \ref{tab:original-dataset-info} presents the statistics of these datasets. They provide citation contexts from various articles where all contexts have a target citation in the middle. The context sizes are in terms of characters, which causes some incomplete words at the start and end of the contexts.

The datasets originally include a "TARGETCIT" marker as a placeholder for citations within each context. We replaced these markers with "$<$mask$>$" tokens to align with our pretraining process. Additionally, to ensure CiteBART focuses solely on predicting target citations, we removed any non-target citations from all four datasets.

We encountered some issues during the preprocessing of ACL-200 and RefSeer. First, they include local contexts with author name conflicts in the citation tokens. For example, the "Petrović et al., 2010" citation token was incorrectly written as "Petrovic et al., 2010" in the target citation column of ACL-200. Another problem is the incorrect ordering of two-author citations. For instance, the local citation context provides the citation "Rivera and Zeinalian, 2016"; the paper metadata includes "Zeinalian and Rivera, 2016". There are also a few cases of incorrect citations. Moreover, there are some contexts with empty author names. We removed all these cases from the aforementioned datasets to ensure consistency.

After the preprocessing, we worked with the train and test sets. As CiteBART involves continual pre-training, we perform it on the training partition and evaluate the performance on the test partition. Table \ref{tab:prerocessed-dataset-info} shows the final statistics of our preprocessed datasets\footnote{Please find information on token limits in Appendix \ref{sec:appendix:token-limits}.} including the training and test partition sizes for all the benchmarks.

% \footnote{Please find information on token limits at \url{https://anonymous.4open.science/r/CitationRecommendation-5FA1/README.md}.}

\begin{table}[ht!]
 \caption{Statistics of the preprocessed datasets.}
  \centering
  \scalebox{0.99}{
  \begin{tabular}{lcccc}
  % \begin{tabular}{p{36mm}p{14mm}p{12mm}p{13mm}l}
    \toprule
    Dataset Name & ACL-200 & PeerRead & RefSeer & Arxiv \\
    \midrule
    \# of local contexts & 63,365 & 16,669 & 3,739,189 & 3,205,210\\
    Size of the training split & 50,692 & 13,335 & 2,991,351 & 2,564,168\\
    Size of the test split & 12,673 & 3,334 & 747,838 & 641,042\\
    \# of removed contexts & 403 & 0 & 39,577 & 0\\
    \# of unique citations & 5,266 & 2,043 & 351,896 & 368,284\\
    \bottomrule
  \end{tabular}}
  \label{tab:prerocessed-dataset-info}
\end{table}

\subsection{Metric Definitions}
\label{sec:metrics}

To evaluate CiteBART, we used the Recall@$10$, Exact Match and Mean Reciprocal Rank metrics. The past works on citation recommendation have generally used Recall@$10$ and Mean Reciprocal Rank as evaluation metrics.

Recall@$10$ is the ratio of the correctly predicted items in the top k recommendations. The benchmark datasets have only one actual target for each context. Therefore, recall@$10$ measures whether the target citation matches any recommendations in top k.

Exact match (EM) calculates whether the first prediction of the model is the same as the target citation. It is the same as accuracy since there is only one ground-truth citation for each context.

Mean Reciprocal Rank (MRR) considers the position of the ground-truth label in a top-k ranked recommendation list. It is the mean of the reciprocal rank of the correctly recommended citation in the recommendation list. Thus, in Equation \ref{eq:1}, $U$ corresponds to the total number of contexts in the dataset (test set size), and $i$ is the position of the ground-truth citation for context $u$ in the top-k results. We used $k$ as $10$ in our experiments.

\begin{equation}
\label{eq:1}
    MRR = \frac{1}{U} \sum_{u=1}^{U}\frac{1}{rank_{i}}
\end{equation}

% *******************************************************************************************************
\section{Experiments}
\label{sec:experiments}

% \footnote{Please find information on training and evaluation times at \url{https://anonymous.4open.science/r/CitationRecommendation-5FA1/README.md}.}

We conducted our experiments on devices with NVIDIA RTX6000 Ada GPU and NVIDIA V100 GPU\footnote{Please find information on training and evaluation times in Appendix \ref{sec:appendix:train-times}.}. The following hyperparameters were utilized in all our experiments. The number of epochs was set to $15$, as the change in loss values between epochs became negligibly small beyond this point. Only the PeerRead Global dataset has been trained for 30 epochs since the generative model requires longer training for the relatively smaller PeerRead dataset. We employed a learning rate of $2e-5$ and an attention dropout rate of $0.12$. Given that BART is a generative model, we adjusted its generation parameters to produce outputs that align with our requirements. Specifically, we utilized the grouped beam search with $20$ beams and applied a diversity penalty of $1.5$ to generate more diverse results. The maximum number of generated tokens was $25$ since the generated citations should not exceed it. Apart from these specific modifications, we did not alter the architecture of the BART model.

\subsection{Results} 

%%% We report our results using Recall@10 (R@10), Exact Match (EM), and Mean Reciprocal Rank (MRR) and compare with the state-of-the-art approaches in Table \ref{tab:comparisons}. As can be seen from the table, CiteBART-Global outperforms others on the existing benchmarks except for the smallest PeerRead dataset, while the base scheme is still a strong baseline, surpassing BERT-GCN on PeerRead, DualEnh, and HAtten on Refseer. The table includes the best-reported results of HAtten with $2k$ pre-fetched candidates. As for DualEnh \citep{medic-snajder-2020-improved}, we chose their superior "DualEnh-ws" model for the comparison. BERT-GCN's \citep{jeong-etal-2020-bertgcn} results are available only on the PeerRead dataset. None of the past works have provided their Exact Match scores.

%%% As shown in Table \ref{tab:comparisons}, CiteBART-Global demonstrates its advantage over HAtten on Refseer most since Refseer includes more training contexts compared to ArXiv. Given that CiteBART is a generative model, access to a larger training set contributes to its improved results.

We report our results using Recall@10 (R@10) and Mean Reciprocal Rank (MRR) and compare with the state-of-the-art approaches in Table \ref{tab:comparisons}\footnote{We share our Exact Match (EM) scores in Appendix \ref{sec:appendix:em-scores}.}. As can be seen from the table, CiteBART-Global outperforms others on the existing benchmarks except for the smallest PeerRead dataset, while the base scheme is still a strong baseline.%, surpassing BERT-GCN on PeerRead, DualEnh, and HAtten on Refseer. 

%  The table includes the best-reported results of HAtten with $2k$ pre-fetched candidates. 
HAtten reports its results based on a $10k$ subset of the test set due to long evaluation times. In Table \ref{tab:comparisons}, however, we present the results of HAtten on the entire test sets.
As for DualEnh \citep{medic-snajder-2020-improved}, we chose their superior "DualEnh-ws" model for the comparison. BERT-GCN's \citep{jeong-etal-2020-bertgcn} results are available only on the PeerRead dataset. We also compare our approach with SymTax \citep{goyal-etal-2024-symtax}; its results surpass Hatten. Additionally, we add BM25 \citep{robertson-bm25}, a fast, TF-IDF-based retrieval function, as a baseline.

As shown in Table \ref{tab:comparisons}, CiteBART-Global demonstrates its advantage over SymTax and HAtten on Refseer most since Refseer includes more training contexts compared to ArXiv. Given that CiteBART is a generative model, access to a larger training set contributes to its improved results.

\begin{table}[htb]
 \caption{Comparison with state-of-the-art on LCR benchmarks. The best values are shown with \textbf{bold}.}
 \centering
 \scalebox{0.90}{
 \begin{threeparttable}
  
  \begin{tabular}{lcccccccc}
    \toprule
    \multirow{2}{*}{Model} &
      \multicolumn{2}{c}{ACL-200} &
      \multicolumn{2}{c}{PeerRead} &
      \multicolumn{2}{c}{Refseer} &
      \multicolumn{2}{c}{Arxiv} \\
      & {R@10} & {MRR} & {R@10} & {MRR} & {R@10} & {MRR} & {R@10} & {MRR} \\
      \midrule

      BM25 & 0.194 & 0.107 & 0.337 & 0.214 & 0.219 & 0.142 & 0.197 & 0.125 \\
      BERT-GCN\tnote{a} & - & - & 0.529 & 0.418 & - & - & - & - \\
      DualEnh-ws & 0.703 & 0.366 & - & - & 0.534 & 0.280 & - & - \\
      %%  HAtten & 0.633 & -\tnote{c} & \textbf{0.757} & -\tnote{c} & 0.454\tnote{b} & -\tnote{c} & 0.439\tnote{b} & -\tnote{c} \\
      HAtten & 0.499 & 0.242 & 0.579 & 0.289 & 0.339 & 0.155 & 0.329 & 0.122 \\

      SymTax (SciV) & 0.653 & 0.296 & \textbf{0.751} & 0.350 & 0.485 & 0.199 & 0.399 & 0.128 \\
      \midrule
      CiteBART-Base & 0.686 & 0.504 & 0.570 & 0.424 & 0.606 & 0.449 & 0.355 & 0.240\\
      CiteBART-Global & \textbf{0.739} & \textbf{0.513} & 0.669 & \textbf{0.502} & \textbf{0.652} & \textbf{0.479} & \textbf{0.502} & \textbf{0.305} \\

      \bottomrule
  \end{tabular}

  \begin{tablenotes}
        \item[a] BERT-GCN performs evaluation by excluding the papers cited less than five times in each dataset.
        % \item[b] The reported results are based on a $10k$ subset of the test set.
        % \item[c] The authors did not report their MRR scores.
  \end{tablenotes}
  \end{threeparttable}}
  
  \label{tab:comparisons}
\end{table}

%%% To observe the citation prediction capabilities of CiteBART in detail, we present a qualitative analysis in Appendix \ref{sec:appendix:llm-analysis}. 

\subsection{Qualitative Analysis}

To provide insights into the working of CiteBART, we present some top $10$ prediction examples. We analyze four different scenarios shown in Table \ref{tab:top-10-prediction-scenarios}. Since CiteBART is a generative model, it is prone to hallucination. In the examples, the hallucinated predictions are designated with the * symbol. 

%\footnote{\url{https://anonymous.4open.science/r/CitationRecommendation-5FA1/utils/additional_materials/Appendix_A.pdf}} 

\begin{table*}[htb]
 \caption{Four example top-10 citation predictions using CiteBART. Due to space limitations, contexts and abstracts have been abbreviated. The hallucinated predictions are designated with the * symbol. The correct predictions are in \textbf{bold}.}
  \centering
  \scalebox{0.77}{
  \begin{tabular}{cccp{15mm}p{18mm}l}
    \toprule
    \# & Context & Ground Truth & Pretraining Dataset of the Model & Dataset of the Example & Predicted Citations\\
    
    \midrule
    \multirow{10}{*}{1} &
    \multirow{10}{63mm}{... Twitter. Previously, a series of NLP tasks have tried to utilize the social annotations like followers , emoticons  and responses \textbf{<mask>} etc. two kinds of common social labels, i.e., hyper-links and hashtags are leveraged for ...} & \multirow{10}{*}{Hu et al., 2015} & \multirow{10}{15mm}{PeerRead Base} & \multirow{10}{18mm}{PeerRead Base}
    & 1. Shang et al., 2015 \\
    & & & & & 2. Vinyals and Le, 2015 \\
    & & & & & 3. Baqapuri, 2015 \\
    & & & & & 4. Serban et al., 2015 \\
    & & & & & 5. Sordoni et al., 2015 \\
    & & & & & 6. Tan et al., 2015 \\
    & & & & & 7. Tan et al., 2014 \\
    & & & & & 8. Yin and Schutze, 2015 * \\
    & & & & & 9. Dhingra et al., 2016 \\
    & & & & & 10. Tan et al., 2016 \\

    \midrule
    \multirow{10}{*}{2} &
    \multirow{10}{63mm}{... Twitter. Previously, a series of NLP tasks have tried to utilize the social annotations like followers , emoticons  and responses \textbf{<mask>} etc. two kinds of common social labels, i.e., hyper-links and hashtags are leveraged for ... </s> TGSum: Build Tweet Guided Multi-Document Summarization Dataset </s> The development of summarization research has been significantly hampered by the ...} & \multirow{10}{*}{Hu et al., 2015} & \multirow{10}{15mm}{PeerRead Global} & \multirow{10}{18mm}{PeerRead Global}
    & \textbf{1. Hu et al., 2015} \\
    & & & & & 2. Vinyals and Le, 2015 \\
    & & & & & 3. Bing et al., 2015 \\
    & & & & & 4. Tan et al., 2014 \\
    & & & & & 5. Dhingra et al., 2016 \\
    & & & & & 6. Xiao and Cho, 2016 \\
    & & & & & 7. Qu and Hovy, 2016 * \\
    & & & & & 8. Bing et al., 2014 * \\
    & & & & & 9. Lei et al., 2015 \\
    & & & & & 10. Qu and Zuidema, 2015 * \\

    \midrule
     \multirow{10}{*}{3} &
     \multirow{10}{63mm}{... in some latent space. There are many ways to structure G. The DCGAN  \textbf{<mask>}  uses fractionally-strided convolutions to upsample images instead of ... </s> Gang of GANs: Generative Adversarial Networks with Maximum Margin Ranking </s> Traditional generative adversarial networks (GAN) and many of its variants are trained by minimizing the KL or JS-divergence loss ...} & \multirow{10}{*}{Radford et al., 2015} & \multirow{10}{15mm}{ACL-200 Global} & \multirow{10}{18mm}{PeerRead Global}
     & 1. Kalchbrenner et al., 2014 \\
     & & & & & 2. Kalchbrenner and Blunsom, 2013 \\
     & & & & & 3. Sha and Pereira, 2003 \\
     & & & & & 4. Mikheev et al., 2013 * \\
     & & & & & 5. Finkel et al., 2008 \\
     & & & & & 6. Mikheev et al., 1999 \\
     & & & & & 7. Gimpel and Smith, 2012 \\
     & & & & & 8. Kim et al., 2014 \\
     & & & & & 9. Blitzer et al., 2006 \\
     & & & & & 10. Henderson, 2004 \\

     \midrule
     \multirow{10}{*}{4} &
     \multirow{10}{63mm}{... models to autoregressive models and stochastic variations of neural networks. Among them \textbf{<mask>} developed an approach for training a generative model with variational inference by performing ... </s> Learning to Generate Chairs, Tables and Cars with Convolutional Networks </s> We train a generative convolutional neural network which is able to generate images of objects given object type, viewpoint ... } & \multirow{10}{*}{Rezende et al., 2014} & \multirow{10}{15mm}{Arxiv Global} & \multirow{10}{18mm}{PeerRead Global} 
     & 1. \textbf{Rezende et al., 2014} \\
     & & & & & 2. Kusner and Hern'andez-lobato, 2016 \\
     & & & & & 3. Gregor et al., 2015 \\
     & & & & & 4. Mnih and Gregor, 2014 \\
     & & & & & 5. Doersch, 2016 \\
     & & & & & 6. Kusner and Hern'andez-lobato, 2015 * \\
     & & & & & 7. Ioffe and Szegedy, 2015 \\
     & & & & & 8. Lamb et al., 2016 \\
     & & & & & 9. Salimans and Kingma, 2016 \\
     & & & & & 10. Salimans and Knowles, 2012 \\
    
    \bottomrule
  \end{tabular}}
  \label{tab:top-10-prediction-scenarios}
\end{table*}

We first present an example context that is tested on a model pre-trained on the PeerRead Base dataset. It belongs to the test set of PeerRead Base and receives top $10$ citation predictions for the mask. As demonstrated below, the model fails to predict the correct citation in the top $10$ predictions. Actually, the ground-truth citation is the $18th$ entry in the ranked prediction list.

In a deeper analysis of the recommended citations for the first example, we bring up their connections with the ground-truth citation. The ground truth citation, "Hu et al., 2015", focuses on sentence-level semantics using convolutional neural networks (CNNs) with an application in dialogue generation. Similarly, the second prediction, "Vinyals and Le, 2015" leverages the sequential structure of sentences in dialogue systems. The fourth prediction, "Serban et al., 2015", also aims to model the hierarchical structure of sentences (utterances) for building an end-to-end dialogue system. The first prediction, "Shang et al., 2015," is still concerned with capturing sentence connections for a generative motivation. However, the primary reason for its top placement should be related to its experiments on Twitter data since the term Twitter appears in the local citation context. Analogously, the predictions $3, 5, 7$, and $9$ utilize Twitter as the data source. Lastly, the model may have proposed the entries $6$ and $10$ due to their overlaps in authors' names with $7$.

The second example has the same context as the first one, but this time, the citing paper's global information (title and abstract) is attached to it. Moreover, the model pre-trained on the PeerRead Global dataset makes the prediction, returning the ground truth citation in the first index. One can observe that the citations "Vinyals and Le, 2015", "Tan et al., 2015", and "Dhingra et al., 2016" still appear in the top-10 prediction list. There are also some hallucinated responses. The newly recommended "Bing et al., 2015" in the third position is also relevant since it tackles constructing sentences from fine-grained textual units.

% The third example highlights our model's cross-dataset capabilities and zero-shot performance. 
The third example highlights our model's cross-dataset generalization capability. We input a context from the PeerRead Global dataset into a model pre-trained on ACL-200 Global. The model fails to predict the correct citation as it is missing in the training dataset. Its predictions are NLP papers since ACL-200 is an NLP corpus. On the other hand, PeerRead includes both vision and text papers. The ground-truth citation, "Radford et al., 2015," focuses on image classification using CNNs, emphasizing unsupervised learning. Our analysis reveals that multiple predicted citations, among the top ten, are relevant to the ground-truth citation. For example, the papers in predictions 1 and 2 also employ CNNs but with a focus on sentence modeling. The papers from predictions 3 and 5 are about conditional random fields (CRFs). While their primary research areas differ significantly from the ground truth, terms such as 'conditional' and 'random' frequently appear in the ground truth paper. Moreover, the paper in Prediction 7 closely aligns with the ground-truth paper by strongly emphasizing unsupervised learning.

The fourth example emphasizes our model's cross-dataset generalization capability from a different perspective. In this example, a model pre-trained on the Arxiv Global dataset manages to correctly predict the ground truth citation for a context from the PeerRead Global dataset. Upon closer inspection, we observed that this citation exists in both datasets but with different contexts. CiteBART-Global can predict the correct ground truth citation for an unseen context, leveraging another context citing the same reference.

%\textcolor{blue}{We classify a predicted citation as a hallucination if it does not belong to the citation pool of the model's pretraining dataset in our citation generation scenario. However, a citation marked as hallucinated may actually exist outside the dataset's pool. We do not consider this as a weakness of our model, as CiteBART can be further pretrained on additional examples to expand its citation pool. Additionally, the hallucinations marked with an * symbol in cross-dataset examples 3 and 4 are considered hallucinations relative to the pretraining datasets of the models. If hallucinations were evaluated from the perspective of the context's dataset, their number would increase, as the model generates citations based on its pretraining dataset's citation pool. 
%\textcolor{blue}{Further qualitative analysis examples are provided in Section \ref{sec:qualitative-hallucinations}. Lastly, our taxonomy of citation hallucinations and a more detailed analysis of hallucinations can be found in Sections \ref{sec:taxonomy} and \ref{sec:analysis-hallucination}, respectively.}

\subsection{Ablation Study}

We conducted an ablation study to show different components' contributions to the overall results. The analysis was carried out on the ACL-200 dataset. Table \ref{tab:ablation} shows the results for CiteBART with a model pre-trained on the ACL-200 Global dataset in $15$ epochs.

The first three experiments test the contribution of the local context, title, and abstract to the overall performance. First, we remove the local context to see the performance due to the global information-only training (\#1 in Table \ref{tab:ablation}). We discard the title and abstract in the second and third configurations (\#2 and \#3 in Table \ref{tab:ablation}). The results show that excluding the local context brings about a sharp reduction in the performance metrics (a drop from $0.739$ to $0.588$ in Recall@10), confirming its decisive role in generating citations. On the other hand, removals of title or abstract do not lead to a statistically significant decrease in performance.

\begin{table}[htb]
 \caption{Ablation study results on ACL-200 Global dataset under four different configurations. The best values are shown with \textbf{bold}.}
  \centering
  \scalebox{0.89}{
  \begin{tabular}{lp{2.3cm}p{8.4cm}ccc}
    \toprule
    & Approach & Training Input & Recall@10 & EM & MRR \\
    \midrule
     & Base & Context & 0.686 & \textbf{0.422} & 0.504 \\
     & Global & Context + Citing Title \& Abstract & \textbf{0.739} & 0.417 & \textbf{0.513} \\ 
    \midrule
    1 & No context & Citing Title \& Abstract & 0.588 & 0.205 & 0.311 \\ 
    2 & No title & Context + Citing Abstract & 0.731 & 0.415 & 0.509 \\ 
    3 & No abstract & Context + Citing Title & 0.712 & 0.396 & 0.490 \\
    4 & All-including & Context + Citing Title \& Abstract + Cited Title \& Abstract & 0.111 & 0.039 & 0.056 \\
    \bottomrule
  \end{tabular}}
  \label{tab:ablation}
\end{table}

In the fourth ablation study, we further expand the global information with the cited paper's title and abstract during pre-training (\#4 in Table \ref{tab:ablation}). The evaluation stays the same, feeding the local context with the citing paper's title and abstract during inference. Contrary to expectations, adding the ground-truth paper's global information during pre-training does not help; the model falls in its performance. This failure may be explained by the model learning to associate the citation token with the global information of both the citing and cited article in the training phase. However, lacking the cited paper's global information in the test phase confuses the model's predictions.

The previous studies (\citet{medic-snajder-2020-improved}, \citet{gu2022hatten}) utilize an all-including training and inference configuration where citing and cited paper's global information is concatenated with the local citation context. Their pre-fetch and re-ranking pipeline is well-suited to this setup and benefits from it as the inference step also allows incorporating the cited paper's title and abstract, which is not the case in a learning approach like ours'. CiteBART-Global outperforms these models without relying on global information about the cited papers, representing a more ideal scenario for the LCR task.

\subsection{Taxonomy and Measurement of Hallucinated Citations}
\label{sec:taxonomy}

CiteBART, similar to other generative models, is prone to hallucination, occasionally producing citations that do not correspond to any real work. A generated citation is classified as hallucination if it is not present in the citation list of the dataset including the input context. Hallucinations in CiteBART are typically entity-error hallucinations or fabrications. %In this section, we analyze the types of hallucinated citations to better understand the issues and possible causes.

%First, we examine hallucinations \textbf{without reference to ground-truth} citations for the corresponding input contexts. These are classified as follows:

%\begin{itemize}
%    \item \textbf{Hallucinated publication year:}
%    The prediction partially matches an existing citation in the reference list of the dataset where the author part is correct but the date part is hallucinated.
    %Hallucinations in this category involve a year that is not feasible (e.g., outside the range of valid publication years) or a year that, when combined with the generated author name(s), does not correspond to a valid citation.
%    \item \textbf{Hallucinated author names:} CiteBART may generate fabricated author names. Generally, they appear as typos.
%    \item \textbf{Fabricated author list:} In some cases, generated citations include an incorrect aggregated list of authors. The associated year may also be irrelevant to either author.
%    \item \textbf{Incorrect citation format:} Occasionally, the generated citation deviates from the author-date citation style. Examples include the use of symbols like “\&” instead of “and” or improper structuring, such as omitting "et al.," where required while only leaving “.,” symbols behind.
%\end{itemize}

%\subsection{Detailed Analysis of Hallucinated Citations}
%\label{sec:analysis-hallucination}

To measure the degree of hallucinations in LLM-generated responses, \citet{li-etal-2024-dawn-hallucination-rates} propose two metrics, MaHR (macro hallucination rate) and MiHR (micro hallucination rate), respectively. While MaHR calculates the proportion of hallucinatory responses in all the responses (Equation \ref{eq:2}), MiHR gives the average rate of hallucinations within each response (Equation \ref{eq:3}).

\begin{equation}
\label{eq:2}
    MaHR = \frac{Count(hallucinatory\ responses)}{n} 
\end{equation}

\begin{equation}
\label{eq:3}
    MiHR = \frac{1}{n} \sum_{i=1}^{n}\frac{Count(hallucinatory\ facts)}{Count(all\ facts\ in\ r_{i})}
\end{equation}

In LCR, MaHR represents the proportion of hallucinated citations across all generated citations. As the task is evaluated with top-$k$ predictions for each test instance, the total number of responses becomes $k*n$ where $n$ is the number of test instances. Thus, MaHR is the fraction of hallucinated citations among $k*n$ responses (Equation \ref{eq:4}).

MiHR, on the other hand, measures the average hallucination rate in individual contexts. For example, each of $n$ contexts gets top-$k$ predictions and yields its hallucination rate, and MiHR is the average of these individual rates (Equation \ref{eq:5}).

\begin{equation}
\label{eq:4}
    MaHR = \frac{Count(hallucinated\ citations)}{k*n} 
\end{equation}

\begin{equation}
\label{eq:5}
    MiHR = \frac{1}{n} \sum_{i=1}^{n}\frac{Count(hallucinated\ citations\ in\ context_{i})}{k}
\end{equation}

In LCR, as each context gets top-$k$ predictions, the number of facts in each response is fixed with $k$ (the denominator in Equation \ref{eq:5}), which makes MaHR and MiHR produce identical results. 

In addition to MaHR (or MiHR), we propose the following metrics to pinpoint hallucination behavior. Each metric targets a type of hallucination we categorized by examining \textbf{hallucinations versus ground truth} citations for given contexts. 

%We examine hallucinations by \textbf{comparing them with the ground truth} citation for a given context. This analysis provides additional insights into the relationship between the generated output and the expected citation. The respective subcategories are as follows:

%\begin{itemize}
%    \item \textbf{Incorrect year (all-names-GT):} The generated citation fully matches the author(s) in the ground truth citation while failing to match the publication year.
%    \item \textbf{Partially correct author list (one-name-GT):} One of the two author names is correct and the generated year may or may not be correct in these cases.
%    \item \textbf{Correct year with incorrect authors (year-GT):} Some hallucinations match the year of the ground truth citation, even if the author names are incorrect.
%    \item \textbf{Completely incorrect predictions:} In this category, there is no alignment with the ground truth. The errors include:
%    \begin{itemize}
%        \item Hallucinations due to incorrect citation formatting (wrong-format).
%        \item Fabricated author names or non-sensical entries.
%        \item Correctly formatted author date citation entries that do not overlap with the ground truth.
%    \end{itemize}
%\end{itemize}

%Through this categorization, we aim to enhance the understanding of the patterns and mechanisms underlying citation-related hallucinations in CiteBART.

\begin{itemize}
    \item \textbf{Incorrect year (all-names-GT):} The generated citation fully matches the author(s) in the ground truth citation while failing to match the publication year.
    \item \textbf{Partially correct author list (one-name-GT):} One of the two author names is correct, and the generated year may or may not be correct in these cases.
    \item \textbf{Correct year with incorrect authors (year-GT):} Some hallucinations match the year of the ground truth citation, even if the author names are incorrect.
    \item \textbf{wrong-format:} If the generated citation's format does not conform to the parenthetical author-date citation style, it is considered a wrong-format hallucination. These types of hallucinations happen very rarely. 
    \item \textbf{other-hal:} The other types of hallucinations that do not belong to any of the above types belong to this category. There is no overlap with any part of GT in these hallucinations. 
\end{itemize}

Additionally, we term the aggregation of the hallucinations corresponding to partially correct responses MaHR-partial and calculate it using Equation \ref{eq:6}. Lastly, we relate MaHR with MaHR-partial using Equation \ref{eq:7}.

\begin{equation}
\label{eq:6}
    MaHR\mhyphen partial = all\mhyphen names\mhyphen GT + one\mhyphen name\mhyphen GT + year\mhyphen GT
\end{equation}

\begin{equation}
\label{eq:7}
    MaHR = MaHR\mhyphen partial + wrong\mhyphen format + other\mhyphen hal
\end{equation}

Table \ref{tab:hallucination-metrics} presents the results of the hallucination metrics for the CiteBART-Global models. To observe the effect of the $k$ value, we performed each analysis with top-3, top-5, and top-10 generated predictions, respectively. The results conclude that MaHR-partial accounts for almost half of the hallucinations in the top 3 predictions, which implies that when the model is forced to make fewer predictions, its hallucinations do not deviate much from the ground truth. The proportion gradually diminishes in the top-5 and top-10 predictions. Interestingly, on Refseer and Arxiv Global, the incorrect year (all-names-GT) hallucination, which is the closest to the ground truth, decreases with increasing $k$ values. In overall performance, the ACL-200 Global dataset gives the lowest hallucination rates all over the $k$ values. Arxiv Global is the second best, with very close scores to ACL-200 Global.

\begin{table*}[htb]
 \caption{Results for proposed hallucination metrics on Global datasets for top-3, top-5, and top-10 predictions. Metric values are shown as percentages (\%). The best values are shown with \textbf{bold}.}
  \centering
  \scalebox{0.83}{
  \begin{tabular}{lcccccccccccc}
    \toprule
    \multirow{2}{*}{Metrics} &
      \multicolumn{3}{c}{ACL-200} &
      \multicolumn{3}{c}{PeerRead} &
      \multicolumn{3}{c}{Refseer} &
      \multicolumn{3}{c}{Arxiv} \\
      & {Top-3} & {Top-5} & {Top-10} & {Top-3} & {Top-5} & {Top-10} & {Top-3} & {Top-5} & {Top-10} & {Top-3} & {Top-5} & {Top-10} \\
      \midrule
            
      all-names-GT & 0.63 & 0.54 & 0.68 & 0.63 & 0.59 & 0.85 & 1.21 & 1.07 & 1.01 & 1.01 & 0.85 & 0.80 \\

      one-name-GT & 0.24 & 0.29 & 0.44 & 1.31 & 1.06 & 1.21 & 0.63 & 0.75 & 0.82 & 0.43 & 0.56 & 0.65 \\

      year-GT & 1.03 & 1.56 & 2.48 & 1.72 & 3.08 & 5.95 & 0.50 & 0.84 & 1.48 & 0.55 & 0.99 & 1.94 \\

      \midrule
      \midrule
      
      \textbf{MaHR-partial}\tnote{a} & \textbf{1.89} & \textbf{2.39} & 3.60 & 3.66 & 4.73 & 8.01 & 2.34 & 2.66 & \textbf{3.31} & 1.99 & 2.40 & 3.39 \\

      wrong-format & 0.02 & 0.02 & 0.08 & 0.00 & 0.07 & 0.28 & 2.18e-5 & 4.97e-5 & 0.01 & 1.35e-5 & 2.12e-5 & 0.01 \\

      other-hal & 2.20 & 4.00 & 9.02 & 4.36 & 7.26 & 15.02 & 2.94 & 5.25 & 10.05 & 2.66 & 4.74 & 9.95 \\

      \midrule
      \midrule

      \textbf{MaHR\tnote{b}} & \textbf{4.12} & \textbf{6.42} & \textbf{12.69} & 8.02 & 12.06 & 23.31 & 5.28 & 7.91 & 13.37 & 4.64 & 7.14 & 13.35 \\
    
%      fabricated-hal-MaHR & 0.99 & 1.43 & 2.48 & 3.97 & 5.15 & 8.04 & 0.63 & 0.94 & 1.59 & 0.50 & 0.82 & 1.45 \\

%      top-k-match-MaHR & 2.54 & 4.55 & 9.69 & 5.05 & 8.15 & 16.53 & 2.76 & 4.74 & 8.60 & 1.79 & 3.12 & 6.28 \\

%      exact-match-MaHR & 2.40 & 3.81 & 7.08 & 4.76 & 6.94 & 12.42 & 2.46 & 3.97 & 6.58 & 1.48 & 2.33 & 3.95 \\

      \bottomrule
  \end{tabular}}

%    \begin{tablenotes}
%        \item[a] $\textbf{MaHR-partial} = \textit{all-names-GT} + \textit{one-name-GT} + \textit{year-GT} $.
%        \item[b]  $\textbf{MaHR} = \textbf{MaHR-partial} + \textit{wrong-format} + \textit{other-hal} $.
%  \end{tablenotes}

    % \begin{tablenotes}
    %     \item[a] The sum of these metrics are equal to MaHR.
    % \end{tablenotes}
  \label{tab:hallucination-metrics}
\end{table*}

Table \ref{tab:hallucination-metrics-more} reports the values of some extended metrics built upon MaHR:

\begin{itemize}
    %\item \textbf{fabricated-hal-MaHR}: This metric shows the rate of only hallucinations with nonsensical fabricated words among all predictions. 
    \item \textbf{top-k-match-MaHR}: This metric considers hallucinated predictions only when one of the other predictions in the same top-k group matches the ground truth (GT).
    \item \textbf{exact-match-MaHR:} This metric is similar to top-k-match-MaHR but specifically focuses on the cases where the exact match occurs (the first prediction is correct).
\end{itemize}

These metrics approach the problem differently by examining the hallucination tendency when the model can hit the ground truth citation in its top-k predictions. In other words, the research question is whether the model suffers less from the hallucination given the correct prediction in the top-k list (when the model knows the answer). The results confirm this hypothesis as top-k-match-MaHR and exact-match-MaHR are different from MaHR in a statistically significant way with $p<0.001$. Furthermore, Arxiv Global is the best model to mitigate hallucinations when it hits the ground truth, outperforming others in the hallucination rates.

\begin{table*}[!hb]
 \caption{Results for extended MaHR metrics on Global datasets for top-3, top-5, and top-10 predictions. Metric values are shown as percentages (\%). The best values are shown with \textbf{bold}.}
  \centering
  \scalebox{0.86}{
  \begin{tabular}{lcccccccccccc}
    \toprule
    \multirow{2}{*}{Metrics} &
      \multicolumn{3}{c}{ACL-200} &
      \multicolumn{3}{c}{PeerRead} &
      \multicolumn{3}{c}{Refseer} &
      \multicolumn{3}{c}{Arxiv} \\
      & {Top-3} & {Top-5} & {Top-10} & {Top-3} & {Top-5} & {Top-10} & {Top-3} & {Top-5} & {Top-10} & {Top-3} & {Top-5} & {Top-10} \\
      \midrule
                
      %fabricated-hal-MaHR & 0.99 & 1.43 & 2.48 & 3.97 & 5.15 & 8.04 & 0.63 & 0.94 & 1.59 & 0.50 & 0.82 & 1.45 \\

      MaHR & \textbf{4.12} & \textbf{6.42} & \textbf{12.69} & 8.02 & 12.06 & 23.31 & 5.28 & 7.91 & 13.37 & 4.64 & 7.14 & 13.35 \\
 
      top-k-match-MaHR & 2.54 & 4.55 & 9.69 & 5.05 & 8.15 & 16.53 & 2.76 & 4.74 & 8.60 & \textbf{1.79} & \textbf{3.12} & \textbf{6.28} \\

      exact-match-MaHR & 2.40 & 3.81 & 7.08 & 4.76 & 6.94 & 12.42 & 2.46 & 3.97 & 6.58 & \textbf{1.48} & \textbf{2.33} & \textbf{3.95} \\

      \bottomrule
  \end{tabular}}

  \label{tab:hallucination-metrics-more}
\end{table*}

% This metric only considers hallucinated predictions when one of the other predictions in their same top-k prediction group match GT. We mainly aim to observe effect of correct prediction of GT on other citations in the same group and whether it increases hallucination rates under these circumstances.       exact-match-MaHR: This metric is similar to top-k-match-MaHR. However, it considers cases with exact match at the first prediction rather than matching any one of the top-k predictions.

\subsection{Qualitative Analysis on Hallucinations}
\label{sec:qualitative-hallucinations}

In this section, we provide additional examples to illustrate the types of hallucinations (Table \ref{tab:additional-qualitative-analysis}). The first example shows an ideal scenario with no hallucinations in the top-10 prediction list. The other examples, except the last, depict different types of hallucinations. The last example showcases the cross-dataset generalization capability of CiteBART. Due to space limitations, contexts and abstracts have been abbreviated. Hallucinated predictions are designated with the * symbol.

\begin{table*}[!hb]
 \caption{Examples of hallucination categories. The referred predictions are in red. \textbf{(a)} No hallucination in any of the top-10 predictions. \textbf{(b)} Hallucinated publication years in the fourth, sixth, and ninth predictions. \textbf{(c)} Hallucinated author name in the sixth prediction. Fabricated author list in the ninth prediction. \textbf{(d)} Hallucinated author name in the fifth prediction. (A typo in the first author's name). \textbf{(e)} Hallucinated author name in the sixth prediction (A single letter as the first author name). \textbf{(f)} CiteBART predicts a citation that has the same author name as the ground truth while in a different citation format and publication year. Unlike the other examples, the model's pretraining dataset is different from the dataset associated with the given context.}
  \centering
  \scalebox{0.75}{
  \begin{tabular}{ccp{25mm}p{15mm}p{18mm}l}
    \toprule
     & Context & Ground Truth & Pretraining Dataset of the Model & Dataset of the Example & Predicted Citations\\

     \midrule
    \multirow{10}{*}{\textbf{(a)}} &
    \multirow{10}{67mm}{... exploits similarity on the target side in another language by extracting source phrases that share common translations \textbf{<mask>} , but recent approaches have combined this approach with source phrase ... </s> Example-based Paraphrasing for Improved Phrase-Based Statistical Machine Translation </s> In this article, an original view on how to improve phrase translation estimates is proposed. This proposal is ...} & \multirow{10}{25mm}{Bannard and Callison-Burch, 2005} & \multirow{10}{15mm}{ACL-200 Global} & \multirow{10}{18mm}{ACL-200 Global}
    & 1. Callison-Burch et al., 2006 \\
    & & & & & 2. Koehn et al., 2003 \\
    & & & & & 3. Irvine and Callison-Burch, 2014 \\
    & & & & & 4. \textbf{Bannard and Callison-Burch, 2005} \\
    & & & & & 5. Quirk et al., 2004 \\
    & & & & & 6. Mirkin et al., 2009 \\
    & & & & & 7. Irvine and Callison-Burch, 2013 \\
    & & & & & 8. Koehn and Knight, 2002 \\
    & & & & & 9. Schroeder et al., 2009 \\
    & & & & & 10. Koehn and Knight, 2003 \\
    
    \midrule
    \multirow{10}{*}{\textbf{(b)}} &
    \multirow{10}{67mm}{... supertags, the supertagger re-analyses the sentence with a more relaxed beam (adaptive supertagging). A* Parsing \textbf{<mask>} a) introduce A* parsing for PCFGs. The parser maintains a chart and an agenda, which is a priority queue of ... </s> A* CCG Parsing with a Supertag-factored Model </s> We introduce a new CCG parsing model which is factored on lexical category assignments. Parsing is then simply a deterministic ...} & \multirow{10}{25mm}{Klein and Manning, 2003} & \multirow{10}{15mm}{ACL-200 Global} & \multirow{10}{18mm}{ACL-200 Global}
    & 1. \textbf{Klein and Manning, 2003} \\
    & & & & & 2. Auli and Lopez, 2011 \\
    & & & & & 3. Ait-Mokhtar and Chanod, 1997 \\
    & & & & & 4. \textcolor{red}{Ait-Mokhtar and Chanod, 2005 *} \\
    & & & & & 5. Pauls et al., 2009 \\
    & & & & & 6. \textcolor{red}{Pauls et al., 2006 *} \\
    & & & & & 7. \textcolor{red}{Ait-Mokhtar and Chanod, 2006 *} \\
    & & & & & 8. Och, 2003 \\
    & & & & & 9. \textcolor{red}{Aitouni et al., 2006 *} \\
    & & & & & 10. Clark and Curran, 2004 \\

    \midrule
    \multirow{10}{*}{\textbf{(c)}} &
    \multirow{10}{67mm}{... Google Analogy Test Set, which contains 14 types of relations with a varying number of instances per relation \textbf{<mask>}, the gger Analogy Test Set , which contains 40 relations with 50 instances per relation, and the ffVec Test Set ... </s> Probabilistic Relation Induction in Vector Space Embeddings </s> Word embeddings have been found to capture a surprisingly rich amount of syntactic and semantic knowledge. However, it is not ...} & \multirow{10}{25mm}{Mikolov et al., 2013} & \multirow{10}{15mm}{PeerRead Global} & \multirow{10}{18mm}{PeerRead Global}
    & 1. Vylomova et al., 2015 \\
    & & & & & 2. Valenzuela-esc\'arcega et al., 2015 \\
    & & & & & 3. Abadi et al., 2016 \\
    & & & & & 4. Heinsohn, 2013 \\
    & & & & & 5. Holzmann and Risse, 2017 \\
    & & & & & 6. \textcolor{red}{Valenzuela-esc\'ar\'arcega et al., 2015 *} \\
    & & & & & 7. Davies et al., 2015 * \\
    & & & & & 8. Dinu et al., 2014 \\
    & & & & & 9. \textcolor{red}{Holzmann and Riedl, 2016 *} \\
    & & & & & 10. Gaunt et al., 2016 \\

    \midrule
    \multirow{10}{*}{\textbf{(d)}} &
    \multirow{10}{67mm}{... produces a false positive rate of 0.0027, as noted above, but in a situation where 3 key-value items were being stored per n-gram on average, this error rate would in fact require a storage cost of 60 bits per original n-gram. 2.2.2 Bloomier Filters More recently, \textbf{<mask>} have proposed an approach to storing large language models which is based on the Bloomier Filter technique of OTHERCIT. Bloomier Filters generalize the Bloom Filter to allow values ...} & \multirow{10}{25mm}{Talbot and Brants, 2008} & \multirow{10}{15mm}{ACL-200 Base} & \multirow{10}{18mm}{ACL-200 Base}
    & 1. \textbf{Talbot and Brants, 2008} \\
    & & & & & 2. Talbot and Osborne, 2007 \\
    & & & & & 3. Lavoie and Rambow, 1997 \\
    & & & & & 4. Pennacchiotti and Pantel, 2009 \\
    & & & & & 5. \textcolor{red}{MTalbot and Brants, 2008 *} \\
    & & & & & 6. Galanis and Androutsopoulos, 2010 \\
    & & & & & 7. Lavoie and Rambow, 2009 * \\
    & & & & & 8. Pennacchiotti and Pantel, 2006 \\
    & & & & & 9. Mintz et al., 2009 \\
    & & & & & 10. Talbot et al., 2011 \\

    \midrule
    \multirow{10}{*}{\textbf{(e)}} &
    \multirow{10}{67mm}{... signature generators can be mislead into generating bad signatures; specifically higher false negative rates. Shield \textbf{<mask>} , Vigilante, DACODA, and our own work, all attempt to work around such problems by directly deriving ... </s> A lightweight end-to-end system for defending against fast worms </s> The vulnerabilities which plague computers cause endless grief to users. Slammer compromised millions of hosts in minutes; a hit-list worm ...} & \multirow{10}{25mm}{Wang et al., 2004} & \multirow{10}{15mm}{Refseer Global} & \multirow{10}{18mm}{Refseer Global}
    & 1. \textbf{Wang et al., 2004} \\
    & & & & & 2. Cui et al., 2007 \\
    & & & & & 3. Brumley et al., 2006 \\
    & & & & & 4. Brumley et al., 2004 * \\
    & & & & & 5. Dasgupta et al., 2004 \\
    & & & & & 6. \textcolor{red}{W et al., 2004 *} \\
    & & & & & 7. Shavitt and Tankel, 2003 \\
    & & & & & 8. Shavitt and Tanenbaum, 2005 * \\
    & & & & & 9. Daswani and S, 2007 * \\
    & & & & & 10. Chen and Wagner, 2007 \\

    \midrule
    \multirow{10}{*}{\textbf{(f)}} &
    \multirow{10}{67mm}{... tab while waiting for the original one to load, i.e., tab switching. More recently, a Web navigation study by \textbf{<mask>} found their participants using multiple windows frequently, enabling them to compare search results ... </s> Parallel Browsing Behavior on the Web </s> Parallel browsing describes a behavior where users visit Web pages in multiple concurrent threads. Web browsers explicitly support this by providing tabs. Although parallel browsing ...} & \multirow{10}{25mm}{Weinreich, 2006} & \multirow{10}{15mm}{ACL-200 Global} & \multirow{10}{18mm}{Refseer Global}
    & 1. \textcolor{red}{Weinreich et al., 2008} \\
    & & & & & 2. Nakagawa and Uchimoto, 2007 * \\
    & & & & & 3. Weinreich et al., 2010 * \\
    & & & & & 4. Navigli and Crisafulli, 2010 \\
    & & & & & 5. Nakashole et al., 2012 * \\
    & & & & & 6. Webber et al., 2003 * \\
    & & & & & 7. Lin and Bilmes, 2011 \\
    & & & & & 8. Resnik and Smith, 2003 \\
    & & & & & 9. Stoica and Hearst, 2004 * \\
    & & & & & 10. Lin and Bilmes, 2008 * \\

    \bottomrule
  \end{tabular}}
  \label{tab:additional-qualitative-analysis}
\end{table*}

\subsection{LLMs in LCR}

LLMs in LCR face a challenge retrieving the top $10$ citations for a given masked context. The main obstacle is the number of candidate citations in the citation pool, which contains $2043$ candidates, even for the smallest PeerRead. It is impractical for an LLM to evaluate every possible citation within a single prompt. Thus, the maximum context length and the size of the citation pool impose a significant bottleneck when applying LLMs to LCR.

To mitigate this issue, \citet{jiang-et-al-2025-llm-analysis-idea} proposed pre-fetching the top 100 candidates using a fast retrieval method such as BM25, and then passing only those candidates to the LLM prompt. Their experiments on the ArXiv and RefSeer datasets reported substantially lower $Recall@10$ scores ($0.134$ and $0.152$, respectively) than CiteBART. Their implementation presents each candidate in a separate prompt and asks for a similarity score in the range ($0-100$) between the ground-truth and candidate citation to reach the overall ranking. As the approach requires $100$ separate prompts per example, the evaluation is prohibitively slow, and the produced similarity score in each case is not directly comparable to those of the others (many repetitive scores), lacking a sufficient basis for the final ranking.

Alternatively, we designed a prompt that simultaneously presented all $100$ pre-fetched citations and asked the LLM to select the top $10$. In practice, however, fitting citation metadata (titles and abstracts) into a single prompt often exceeded context length limits, and even when feasible, models frequently failed to select citations, producing invalid outputs. We also tested a simplified version, asking the LLM to return only the best citation for the exact match evaluation. Although this worked occasionally, the model often defaulted to echoing the top-ranked BM25 candidate. Our results suggest that the LCR task is currently quite challenging for LLMs due to prompt design and efficiency bottlenecks. We provide a qualitative analysis on the performance of LLMs in LCR in Appendix \ref{sec:appendix:llm-analysis}.

% *********************************************************************************************************

\section{Discussion and Conclusion}

%This work proposes CiteBART, a custom language model pre-training with citation objectives for LCR. In CiteBART-Base, we mask the citation tokens in the local contexts to make citation predictions. This local context-only pre-training performs as a good baseline, superior to BERT-GCN on PeerRead, and DualEnh-ws and HAtten on Refseer. CiteBART-Global concatenates the title and abstract of the citing paper to the local context during the citation-masked pre-training. Its superior performance proves the effectiveness of global information, such as titles and abstracts, in the citation learning task.

CiteBART is distinctive as it performs LCR by end-to-end learning. On the other hand, the recent approaches adopt pre-fetch and re-rank pipelines where their system first retrieves a set of papers and then ranks the retrieved by matching queries (citing papers' titles and abstracts, local citation contexts) with candidate papers' representations (cited papers' titles and abstracts). While our model does not use global information about cited papers during testing, these systems require titles and abstracts of the cited papers for inference. In CiteBART-Base, we rely solely on local citation contexts, while CiteBART-Global incorporates the citing paper's global information to make predictions. CiteBART-Global achieves state-of-the-art performance on LCR benchmarks except for the FullTextPeerRead dataset, which is quite small to see the advantage of generative pre-training.

CiteBART can still be fine-tuned for any downstream task. We hypothesize that it should perform better in downstream tasks involving citations and scientific papers than other language models without citation-specific learning signals during pre-training, an area we intend to explore in future work. Furthermore, with the release of new citation recommendation datasets, it will be sufficient to continually pre-train the model to acquire knowledge about the new scientific papers with no need to pre-train from scratch.

We comment on the pros of using BART over encoder-based pre-training models such as RoBERTa. BART's MLM objective is flexible and allows the masking of all the tokens in the parenthetical author-date style. RoBERTa cannot add citation tokens to its vocabulary by its MLM. Moreover, constraining predictions to citation tokens for RoBERTa is not straightforward. While BART is prone to hallucination, its capabilities significantly enhance LCR performance.

Furthermore, our comprehensive hallucination analysis sheds light on the hallucination behavior, MaHR-partial taking up significant proportions (almost half of the hallucinations in the top 3 predictions), which implies that all the hallucinations should not be rejected beforehand but show signs of promising zero-shot capabilities as MaHR-partial is the aggregation of partially correct hallucinations that are correct in all the author names, single author names, and year, respectively. The hallucinations that are (partially) correct in the author names may be useful for finding suggested reading material along with the ground truth paper as they reveal relevant authors. Another finding is that when the prediction is successful in the top-k list, the hallucination tendency in the other predictions drops significantly, the Arxiv Global trained model being the most advantageous, highlighting that the largest model also shows good traits in mitigating hallucinations. The evidence on hallucinations in this study may also lead to hallucination analyses in other domains that clear up generative models' hallucination landscape. 

As shown in our ablation study, extending the local citation context with both the citing and cited paper's title and abstract during the continual pre-training does not produce a better result, which can be evaluated counter-intuitive as one has all the information to learn a citation relationship. 
The missing global information for the cited paper in the test phase complicates finding out the associated citation token.

For future work, we plan to investigate further the all-including configuration given in the ablation study. Conceptually, exploiting the cited paper’s title and abstract during the continual pre-training should have been complementary. However, the empirical evidence proves the contrary. More sophisticated masking strategies besides citation token masking should connect the dots by combining the information from the citing paper’s title and abstract, local citation context, and the cited paper’s title and abstract.
We also plan to investigate the connection between custom mask filling and the recognition of retrieval tokens in the context of generative information retrieval methods.We believe it is feasible to integrate custom citation mask-filling mechanisms with text generation models capable of producing citation placeholders.
Additionally, we should investigate the potential solutions to the citation-specific hallucinations and tackle a way to reduce the number of hallucinated recommendations in the top k. %Force alignment can be considered to remove hallucinations to turn the solution into an end product.

% ***********************************************************************************************************
\section*{Limitations}

We recognize the following limitations in this study. First, CiteBART addresses the task of LCR, predicting the best candidates for a citation placeholder in a given context. As a citation placeholder indicates that the context is worth citation, CiteBART builds upon the assumption of the citation worthiness of a local context.

Second, CiteBART necessitates pre-training on a specific dataset to recommend citations from the pool of papers in it. Thus, it may omit to cite some work or authors if they are not included in its training corpus. However, unlike the past works, as CiteBART is generative, it can recommend unseen papers, hallucinating. Although the fabricated citations in the top k predictions show that they capture the author names of the ground-truth citations, hallucination is still a problem.

Moreover, extending CiteBART to handle multi-citation scenarios, where a context refers to multiple citations simultaneously, would make the task setting more realistic for LCR. However, the current four LCR benchmarks only provide metadata (title and abstract) for the middle citation in each context, while other citations' metadata are removed. Supporting multi-citation contexts would require minor modifications to our model architecture and codebase. Yet, more importantly, it necessitates constructing an LCR dataset specifically designed to include multiple citations (with all their metadata) per context.

There can be a bias towards citing papers as CiteBART learns from both local context and citing papers. Leveraging all the parts of a citation relationship, citing paper, local context, and cited paper should provide a more balanced learning process once it can be made learning. We leave this possibility for future exploration.

%We recognize the following limitations in this study. First, CiteBART addresses the task of LCR, and given context with a citation placeholder, it predicts the best candidates for the placeholder. As a citation placeholder indicates that the context is worth citation, CiteBART builds upon the assumption of the citation worthiness of a local context.

%Second, CiteBART necessitates pre-training on a specific dataset to recommend citations from the pool of papers in it. Thus, it may omit to cite some work or authors if they are not included in its training corpus. However, unlike the past works, as CiteBART is generative, it can recommend unseen papers, hallucinating. Although the fabricated citations in the top k predictions show that they capture the author names of the ground-truth citations, hallucination is still a problem.

%There can be a bias towards citing papers as CiteBART learns from both local context and citing papers. Leveraging all the parts of a citation relationship, citing paper, local context, and cited paper should provide a more balanced learning process once it can be made learning. We leave this possibility for future exploration.

\section*{Ethics Statement}

CiteBART is a tool to support the scientific community in paper writing; it in no way replaces a researcher or alternates the thoughtful process of choosing the most appropriate references to cite in a local context.

\section*{Acknowledgments}
The Scientific and Technological Research Council of Türkiye (TUBITAK) supported this research with the 2219 fellowship awarded to Selma Tekir as a visiting scholar at the University of Edinburgh School of Informatics. Selma is grateful to Mark Steedman for his hospitality and their fruitful discussions.

We primarily used the hardware purchased by the project supported by the Council of Higher Education (YÖK) under ADEP grant number 2022IYTE-3-0027 for our experiments. They were partially run at TUBITAK ULAKBIM, High Performance and Grid Computing Center (TRUBA resources).

%*************************************************************************************************************************

% Bibliography
\bibliographystyle{unsrtnat}  % unsrt
\bibliography{references}

\appendix

\section{Token Limits}
\label{sec:appendix:token-limits}

Before pre-training with citation objectives, we ensured that each context has its "$<$mask$>$" token in its middle position after tokenization. Another critical aspect was the determination of correct lengths for citation contexts. We limited citation contexts in each dataset to an optimal number of tokens to avoid increasing time and memory costs. An exploratory analysis of context lengths shows that the contexts of ACL-200 and Peerread are significantly longer than those of the other datasets. After tokenization, we observed that $200-400$ tokens were optimal for all base datasets. This limit allows sufficiently long contexts without a need for excessive amounts of padding tokens. As an exception, ACL-200 has $607$ contexts that exceed the $400$ limit. We have shortened them to the $400$ token limit as they correspond to a small proportion of the whole number of contexts and also because the number of discarded tokens is negligible. 

\begin{table*}[ht]
 \caption{Maximum token limits for the preprocessed datasets.}
  \centering
  \scalebox{0.94}{
  \begin{tabular}{lll}
    \toprule
    Dataset Name & Base Token Limit & Global Token Limit \\
    \midrule
    ACL-200 & 400 & 350 \\
    FullTextPeerRead & 400 & 350 \\
    Refseer & 200 & 350 \\
    Arxiv & 300 & 350 \\
    \bottomrule
  \end{tabular}}
  \label{tab:max-token-limits}
\end{table*}

For each global dataset, we chose the token limit as $350$. Since abstracts require a higher number of tokens, we limited the local context sizes to $100$ for the global versions of the datasets. We also ensured that there are $50$ tokens each on the left and right sides of the <mask> tokens. We used a token limit of $200$ for abstracts for all datasets since most abstracts can fit into it. Table \ref{tab:max-token-limits} shows the maximum token limits for both the base and global training schemes.

\section{Training and Evaluation Times}
\label{sec:appendix:train-times}

We conducted our experiments on devices with NVIDIA RTX6000 Ada GPU and NVIDIA V100 GPU for Global and Base datasets, respectively. For global datasets, the pre-training for Peerread and ACL-200 lasts for $2$ and $6$ hours, respectively. The larger datasets, Arxiv and Refseer, take up to $8-9$ days since they have similar sizes. For base datasets, the training for the smaller datasets, Peerread and ACL-200, lasts for 8 and 20 hours, respectively. The larger datasets, Arxiv and Refseer, take up to 14-15 days. However, we believe these relatively longer times are the result of training on the device with NVIDIA V100 GPU.

Our evaluation of the corresponding test sets takes considerable time since generating the top 10 predictions for each example is resource-intensive. Especially with our limited hardware resources, acquiring the results on the larger datasets takes up to 2 days. The smaller datasets require less time, 20 minutes for Peerread and 2 hours for ACL-200. We performed our evaluations on the device with NVIDIA RTX6000 Ada GPU.

The issue of slow evaluation for larger datasets is not exclusive to our work. \citet{gu2022hatten} reported their results using only a smaller subsection ($10K$) of the test sets due to long evaluation times.

\section{Exact Match Scores}
\label{sec:appendix:em-scores}

Table \ref{tab:em-scores} presents the exact match (EM) scores of CiteBART. While previous studies did not report EM scores, we consider this metric valuable for assessing the model’s ability to generate the correct citation on its first attempt. As shown in the table, CiteBART successfully predicts the correct citation directly for a substantial portion of the benchmark datasets.

\begin{table}[htb]
 \caption{Exact Match (EM) score of CiteBART on LCR benchmarks.}
 \centering
 \scalebox{0.74}{
  \begin{tabular}{lcccc}
    \toprule
    \multirow{2}{*}{Model} &
      \multicolumn{1}{c}{ACL-200} &
      \multicolumn{1}{c}{PeerRead} &
      \multicolumn{1}{c}{Refseer} &
      \multicolumn{1}{c}{Arxiv} \\
       & {EM} & {EM} & {EM} & {EM} \\
      \midrule
      
      CiteBART-Base & \textbf{0.422} & 0.363 & 0.382 & 0.184 \\
      CiteBART-Global & 0.417 & \textbf{0.430} & \textbf{0.404} & \textbf{0.230} \\

      \bottomrule
  \end{tabular}}
  
  \label{tab:em-scores}
\end{table}

\section{Qualitative Analysis on Large Language Models' Performances in LCR}
\label{sec:appendix:llm-analysis}

We conducted experiments on a Large Language Model (LLM) to evaluate its performance in local citation recommendation. We prompted the open-source "Llama-2-70b-chat" model for our trials. In each prompt, we first list a set of citation tokens ($200$, due to the limits of chat windows) from our dataset, followed by a few examples of masked contexts with the corresponding ground truth mask values. Subsequently, we ask the model to fill in the mask for a new context by selecting a citation from the initially provided list.

\begin{figure*}[ht]
    \centering
    \includegraphics[width=0.79\linewidth]{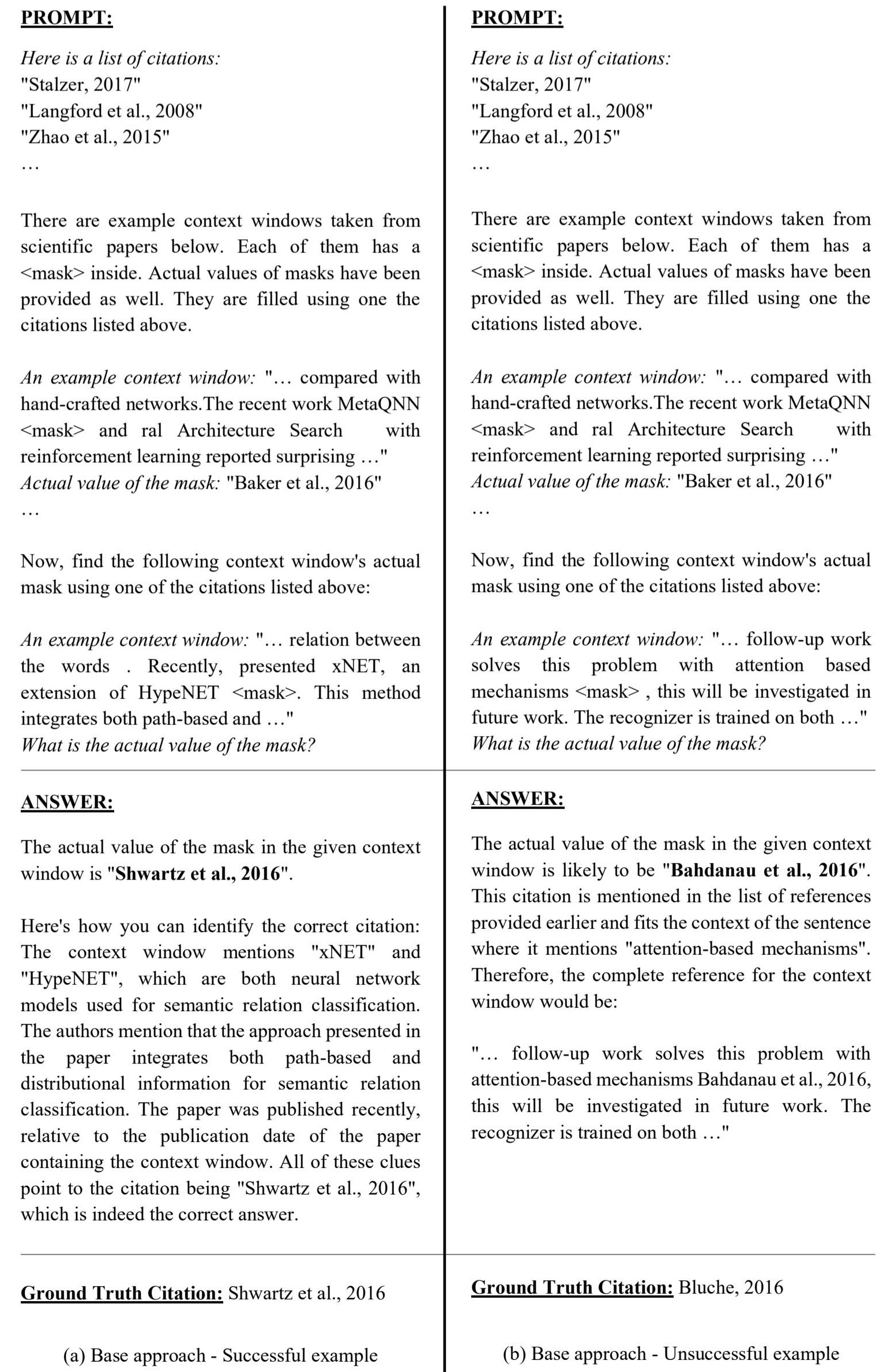}
    \caption{Prompt examples on a Large Language Model for Base dataset.}
    \label{fig:llm-prompting-trial-base}
\end{figure*}

\begin{figure*}[ht]
   \centering
   \includegraphics[width=0.80\linewidth]{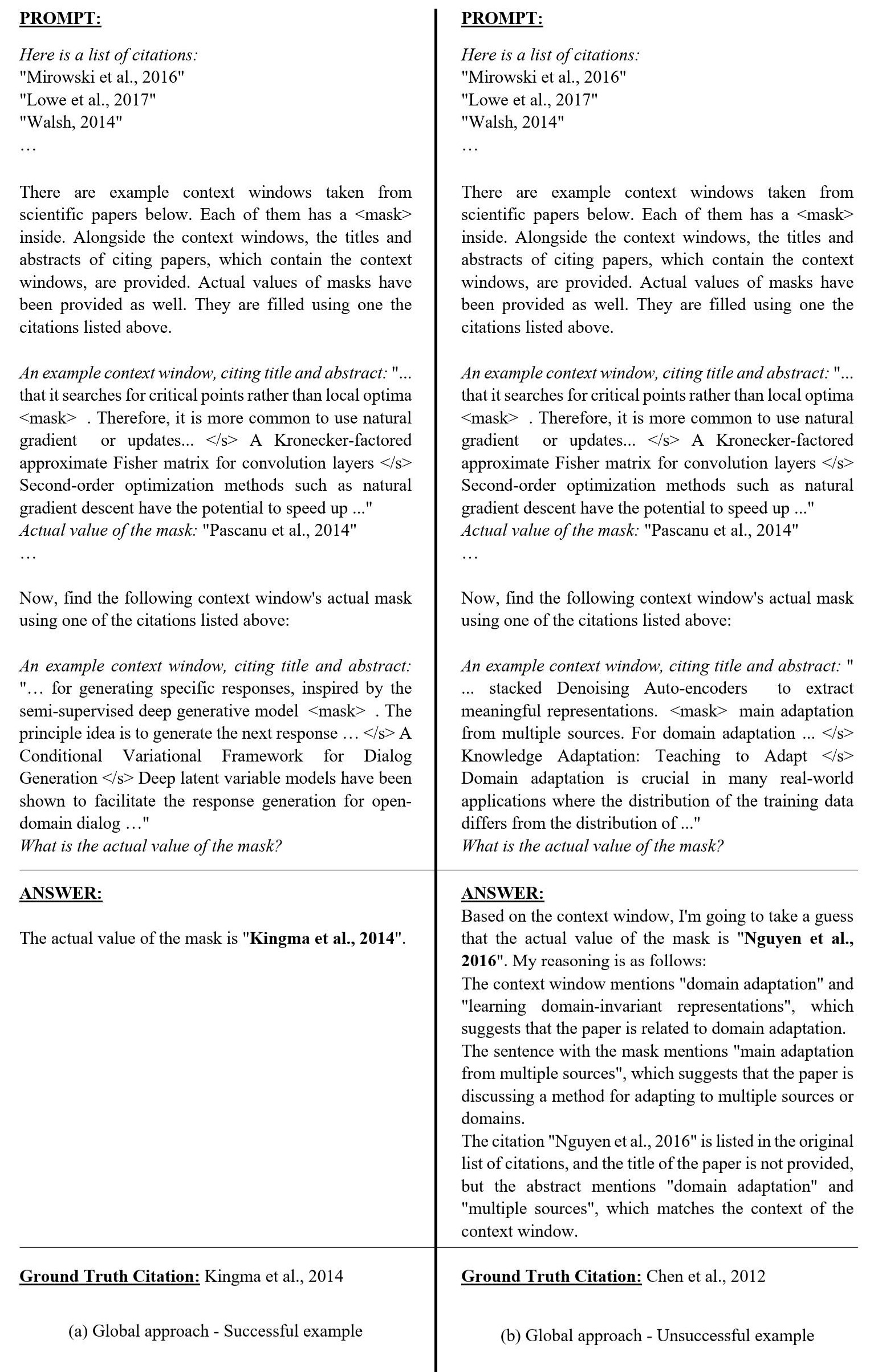}
   \caption{Prompt examples on a Large Language Model for Global dataset.}
   \label{fig:llm-prompting-trial-global}
\end{figure*}

We present four examples in Figures \ref{fig:llm-prompting-trial-base} and \ref{fig:llm-prompting-trial-global} to illustrate the workings of the base and global pre-training schemes, respectively. Due to space constraints, we partially display the list of citations, example contexts, and citing abstracts in the prompts. Each example consists of three parts: the prompt, the LLM's answer, and the ground truth value of the masked citation token provided at the end of the prompt.

Figure \ref{fig:llm-prompting-trial-base} includes a correct prediction in Part (a) and an incorrect one in (b). Indeed, the correct prediction is the only successful example in several trials using the base approach. The model responds to the prompt by "Shwartz et al., 2016" explaining its choice. On the other hand, the model fills in the mask by "Bahdanau et al., 2016" in Part (b), where "Bluche, 2016" is expected. Its reasoning sheds light on its wrong choice as it strongly associates the term "attention-based mechanisms" in the local context with Bahdanau et al.'s seminal paper on attention-based sequence modeling.

In Figure \ref{fig:llm-prompting-trial-global}, Part (a) presents a successful example based on the global dataset where the prompt includes the citing paper's title and abstract with the local citation context. The LLM generates the correct citation without an explanation, unlike other predictions. The second example in Part (b) belongs to an incorrect prediction, yet the LLM makes a plausible choice here, judging from its grounding. We can conclude from the observed behavior that LLMs need custom pre-training for the citation tokens to perform well in the task of local citation recommendation. 

Our further trials with LLMs demonstrate that they tend not to restrict their predictions to the provided list of citations but to recommend the best choice based on their prior knowledge. They also exhibit a known deficiency. They sometimes ask for confirmation when they provide an answer, and even if you confirm, they lean towards changing the answer. In conclusion, they suffer from hallucinations.

\end{document}